\documentclass{article}

\usepackage{microtype}
\usepackage{graphicx}
\usepackage{subfigure}
\usepackage{booktabs} 
\usepackage{multirow} 
\usepackage{amsmath} 

\usepackage{hyperref}
\usepackage{enumitem}

\usepackage[utf8]{inputenc}
\usepackage[T1]{fontenc}
\usepackage{tabularx}
\usepackage{booktabs}
\usepackage{xcolor} 
\usepackage{caption}

\definecolor{highlightRed}{RGB}{255, 230, 230}
\definecolor{highlightBlue}{RGB}{230, 242, 255}

\usepackage[preprint]{icml2025}

\usepackage{amsmath}
\usepackage{amssymb}
\usepackage{mathtools}
\usepackage{amsthm}

\usepackage[capitalize,noabbrev]{cleveref}

\theoremstyle{plain}

\theoremstyle{definition}

\theoremstyle{remark}

\usepackage[textsize=tiny]{todonotes}

\newcommand{\method}{\textit{CodeCircuit}}
\icmltitlerunning{\method{}: Toward Inferring LLM-Generated Code Correctness via Attribution Graphs}

\begin{document}

\twocolumn[
\icmltitle{\method{}: Toward Inferring LLM-Generated Code Correctness via Attribution Graphs}



\icmlsetsymbol{equal}{*}

\begin{icmlauthorlist}
\icmlauthor{Yicheng He}{equal,uiuc}
\icmlauthor{Zheng Zhao}{equal,edin}
\icmlauthor{Kaiyu Zhou}{ntu}
\icmlauthor{Bryan Dai}{IQuest}
\icmlauthor{Jie Fu}{IQuest}
\icmlauthor{Yonghui Yang}{nus}
\end{icmlauthorlist}

\icmlaffiliation{uiuc}{University of Illinois Urbana-Champaign}
\icmlaffiliation{edin}{University of Edinburgh}
\icmlaffiliation{ntu}{Nanyang Technological University}
\icmlaffiliation{nus}{National University of Singapore}
\icmlaffiliation{IQuest}{IQuest Research}

\icmlcorrespondingauthor{Yicheng He}{yh84@uiuc.edu}

\icmlkeywords{Machine Learning, ICML}

\vskip 0.3in
]



\printAffiliationsAndNotice{\icmlEqualContribution} 

\begin{abstract}
Current paradigms for code verification rely heavily on external mechanisms—such as execution-based unit tests or auxiliary LLM judges—which are often labor-intensive or limited by the judging model's own capabilities. This raises a fundamental, yet unexplored question: Can an LLM's functional correctness be assessed purely from its internal computational structure? Our primary objective is to investigate whether the model’s neural dynamics encode internally decodable signals that are predictive of logical validity during code generation. Inspired by mechanistic interpretability, we propose to treat code verification as a mechanistic diagnostic task, mapping the model's explicit algorithmic trajectory into line-level attribution graphs. By decomposing complex residual flows, we aim to identify the structural signatures that distinguish sound reasoning from logical failure within the model’s internal circuits. Analysis across Python, C++, and Java confirms that intrinsic correctness signals are robust across diverse syntaxes. Topological features from these internal graphs predict correctness more reliably than surface heuristics and enable targeted causal interventions to fix erroneous logic. These findings establish internal introspection as a decodable property for verifying generated code. Our code is at~\url{https://github.com/bruno686/CodeCircuit}.
\end{abstract}

\section{Introduction}
Large language models (LLMs;~\citealp{liu2024deepseek, he2025visplay,Roziere2023CodeLlama,Guo2024DeepSeekCoder}) have transformed code generation from simple solvers to complex agents~\cite{Wang2024OpenDevin,shinn2023reflexion,madaan2023selfrefine}. As these models integrate into mission-critical workflows, the demand for functionally correct and maintainable code has intensified~\cite{liu2023your,liu2024evaluating}. However, the stochastic nature of LLMs makes reliable assessment a persistent bottleneck. While execution-based verification via unit tests remains the standard~\cite{khan2024xcodeeval,dong2025codescore,wang2023execution,chen2021codex_humaneval,austin2021mbpp,hendrycks2021apps}, it is limited by labor-intensive test design and sparse edge-case coverage. Emergent \textit{LLM-as-a-Judge} frameworks offer scalable automation~\cite{llm_asjudge_1,llm_asjudge_2}, yet they introduce inference overhead and suffer from recursive dependencies where reliability is bounded by the evaluator's own reasoning priors.

Given these limitations of external mechanisms, a natural question arises:~\textit{Can code correctness be assessed intrinsically, by examining the LLM's internal computational structure?} Prior work has probed neuron activations and representation dynamics during generation~\cite{bui2025correctness, huang2025risk, patel2025planning}, yet these analyses have not been directly linked to the correctness of the generated code. In parallel, recent advances in mechanistic interpretability~\cite{mc1,mc2,mc3,openai_mc,anthropic_mc}, specifically the use of attribution graphs to trace reasoning pathways~\cite{dunefsky2024transcoders,anthropic_attributiongraph, zhao2025verifying}, have shown promise in reasoning domains. We believe that bridging this gap requires a deeper look at how code-specific structures are manifested within the model's latent space during the generation process.

Unlike natural language or abstract math, code possesses a rigid, objective topology, such as control flow and variable binding. Recent research~\cite{zhang2025fsa_cot} demonstrates that LLMs could form internal finite automata rather than relying on superficial statistical shortcuts. Thus, we hypothesize that reliable code generation requires the model to instantiate specific neural circuits that mirror this topology, and that errors manifest as detectable structural collapses within these circuits. By decoding the intrinsic circuits that drive code generation, we gain a window into the model's internally consistent computational trajectories—bypassing both the opacity of black-box methods and the incompleteness of external verification frameworks, ultimately providing a more robust foundation for assessing model reliability.

\begin{figure*}[th!]
  \centering
  \includegraphics[width=\textwidth]{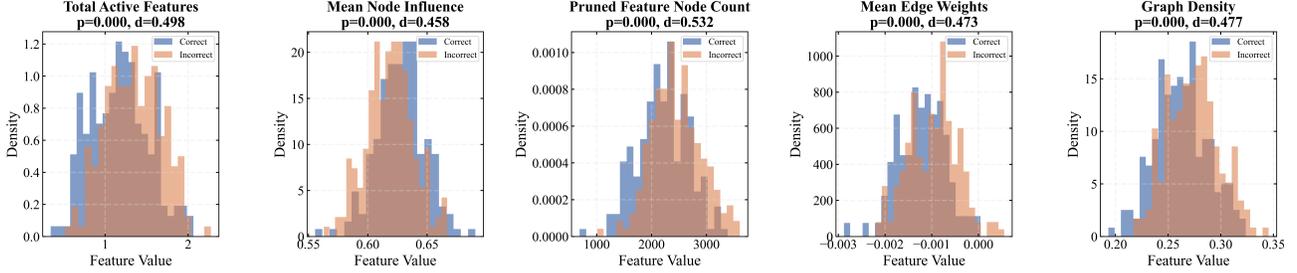}
  \vspace{-15pt}
  \caption{Topological fingerprints of code generation errors. Distributions of five graph features extracted from attribution graphs show differences between correct (blue) and incorrect (orange) code construction steps. These results demonstrate that the attribution topology provides a structural signal for monitoring the integrity of the code generation process.}
  \vspace{-5pt}
  \label{topological}
\end{figure*}

To operationalize this, we introduce~\method{}, a white-box verification framework that analyzes the line-level attribution graph of the generated code. We adopt the instrumentation of sparse autoencoders (transcoders) to decompose complex residual flows into a causal graph of interpretable features. Using~\method{}, we conduct an extensive analysis across diverse programming languages (Python, C++, Java).
Although line-level attribution graphs are computationally demanding, this detail is crucial for exposing failure mechanisms. \method{} thus serves as a \textit{scientific instrument} for precise auditing and debugging, providing mechanistic insights into code generation beyond what coarser methods can reveal.
Our results reveal that, within a fixed model, internal attribution structures correlate with code correctness across multiple programming languages.
We find that correct and incorrect code exhibit systematic topological differences in their attribution graphs (Figure~\ref{topological}). Consequently,~\method{} consistently outperforms both black-box methods (e.g., Temperature Scaling;~\citealp{ts}) and gray-box methods (e.g., Chain-of-Embedding;~\citealp{coe}) in verification performance. Notably, we demonstrate that these signals are not merely correlational: by performing targeted interventions on the graph's nodes~\cite{meng2022rome,meng2023memit}, we can causally correct erroneous code, proving that the attribution graph captures the functional mechanism of generation. Our contributions are summarized as follows:
\begin{itemize}[leftmargin=*] 
\item \textbf{Assessing Generated Code via Internal Computational Structure.} We investigate the feasibility of evaluating LLM-generated code by examining the model's internal neural dynamics rather than its surface-level output. By extracting line-level attribution graphs, we demonstrate that code correctness appears to be reflected in structured patterns of internal computation, enabling reliability assessment without external supervision or execution.
\item \textbf{Shared Signals and Causal Debugging.} We provide empirical evidence that these structural correctness signals are robust across multiple programming languages, suggesting that LLMs learn shared computational circuits independent of surface syntax. Furthermore, we establish the causal nature of these features: by intervening on specific nodes within the attribution graph, we demonstrate the ability to patch the model's internal logic, moving from passive verification to active mechanistic debugging.
\end{itemize}

\section{Preliminary}
\textbf{Mechanistic Interpretability via Attribution Graph.} Attribution Graphs (AGs;~\citealp{anthropic_attributiongraph}) provide a causal, linear decomposition of a transformer model's computation for a specific input prompt, enabling the tracing of information flow through interpretable features. The construction of an AG is predicated on the local replacement model, which substitutes the standard Multi-Layer Perceptrons (MLPs) with a set of Per-Layer Transcoders (PLTs). A PLT creates a sparse feature vector $f^{(l)}$ at layer $l$ from the residual stream $x^{(l)}$ via an encoder $W_{enc}^{(l)}$ and a non-linearity $\sigma$ (e.g., TopK or ReLU):
\begin{equation}
    f^{(l)} = \sigma(W_{enc}^{(l)} x^{(l)} + b_{enc}^{(l)}).
\end{equation}
Crucially, these features are trained to reconstruct the local MLP output (or residual update) via decoder weights $W_{dec}^{(l)}$. Unlike standard Autoencoders, PLTs disentangle the superposition of features in the dense activation space:
\begin{equation}
    \hat{m}^{(l)} = W_{dec}^{(l)} f^{(l)} + b_{dec}^{(l)}.
\end{equation}
The local replacement model is constructed by replacing the true MLP output $m^{(l)}$ with an error-corrected PLT output $\hat{m}^{(l)}_{adj} = \hat{m}^{(l)} + (m^{(l)} - \hat{m}^{(l)})$, where the error term $(m^{(l)} - \hat{m}^{(l)})$ is computed on the forward pass and effectively acts as a bias node. This yields a locally linearized computation in which downstream quantities are linear functions of feature activations and residual stream components, with attention outputs and normalization statistics treated as fixed. Linearity therefore holds with respect to the substituted MLP pathways under the frozen forward pass.

The AG is a directed acyclic graph $G=(V, E)$ where nodes $V$ include active PLT features, token embeddings, error terms, and output logits. An edge $w_{ij} \in E$ from a source node $i$ (at layer $l$) to a target node $j$ (at layer $l' > l$) represents the linear contribution of node $i$'s activation to the pre-activation of node $j$ under the locally linearized computation. Let $a_i$ denote the activation of node $i$, and let $v_{out,i}$ and $v_{in,j}$ denote the residual stream directions associated with the output of node $i$ and the input of node $j$, respectively. The contribution is defined as:
\begin{equation}
    w_{ij} = a_i \, v_{in,j}^\top J_{ij} v_{out,i},
\end{equation}
where $J_{ij}$ is the Jacobian of the frozen residual stream transformation mapping the output of node $i$ to the input of node $j$, with stop-gradients applied to all nonlinearities. This formulation attributes influence through the residual stream rather than via direct feature-to-feature dependencies.

To enhance interpretability, the graph undergoes a pruning process. Rather than matrix inversion (which is intractable for large feature spaces), nodes are pruned based on their attribution to the final model output. We compute the attribution of every feature node to the correct logit (or a target metric) via a single backward pass. Nodes are retained only if they contribute a significant portion of the total influence relative to a threshold $\tau$. Edges with negligible weights are similarly pruned, resulting in a sparse, mechanistically faithful circuit responsible for the model's behavior.

\textbf{Problem Formulation.}
\label{sec:problem_formulation}
We consider the task of verifying the correctness of a code snippet $\mathcal{C}$ generated by a large language model in response to a natural language specification $\mathcal{S}$. Formally, the code snippet is decomposed into a sequence of $N$ discrete logical steps (e.g., lines of code), denoted as $\mathcal{C} = \{s_1, s_2, \dots, s_N\}$. Our objective is to predict the correctness label $y_i \in \{0, 1\}$ for each step $s_i$, where $y_i = 1$ indicates that the $i$-th step is logically sound and consistent with the specification $\mathcal{S}$, and $y_i = 0$ otherwise.

Unlike traditional verification methods that rely on the execution output of the entire snippet $\mathcal{C}$, we formulate code verification as a \textit{mechanistic sequence labeling problem}. For each step $s_i$, the LLM's internal computation is represented by a high-dimensional activation trace. We map this trace onto an interpretable basis to construct a step-level attribution graph $\mathcal{G}_i = (V_i, E_i)$, which serves as a formal proxy for the model's latent algorithmic execution. The problem is thus transformed into learning a diagnostic mapping $F: \mathcal{G}_i \to y_i$.

This mapping $F$ is decomposed into two successive operations. First, a structural encoding function $\Phi$ distills the complex, variable-sized topology of the attribution graph into a fixed-size feature vector $\mathbf{x}_i \in \mathbb{R}^d$:
\begin{equation}
    \mathbf{x}_i = \Phi(\mathcal{G}_i),
\end{equation}
where $\mathbf{x}_i$ captures the structural fingerprints of the computational trajectory, including node composition, information flow density, and topological bottlenecks. Second, a discriminative classifier $h_\phi$, parameterized by $\phi$, maps the structural encoding to a correctness probability:
\begin{equation}
    \hat{y}_i = h_\phi(\mathbf{x}_i) = P(y_i = 1 \mid \mathcal{G}_i; \phi).
\end{equation}
The learning objective is to find the optimal parameters $\phi^*$ that minimize the empirical risk over a dataset of $M$ labeled code steps:
\begin{equation}
    \phi^* = \arg\min_{\phi} \frac{1}{M} \sum_{j=1}^{M} \mathcal{L}(h_\phi(\Phi(\mathcal{G}_j)), y_j),
\end{equation}
where $\mathcal{L}$ is a standard cross-entropy loss function. This formulation enables the detection of logical fallacies at a granular, step-by-step level, grounded in the structural pathologies of the model's internal reasoning pathways.

\section{\method{}: Mechanistic Code Verification}

\label{sec:methodology}

\begin{figure*}[t]
  \centering
  \includegraphics[width=\textwidth]{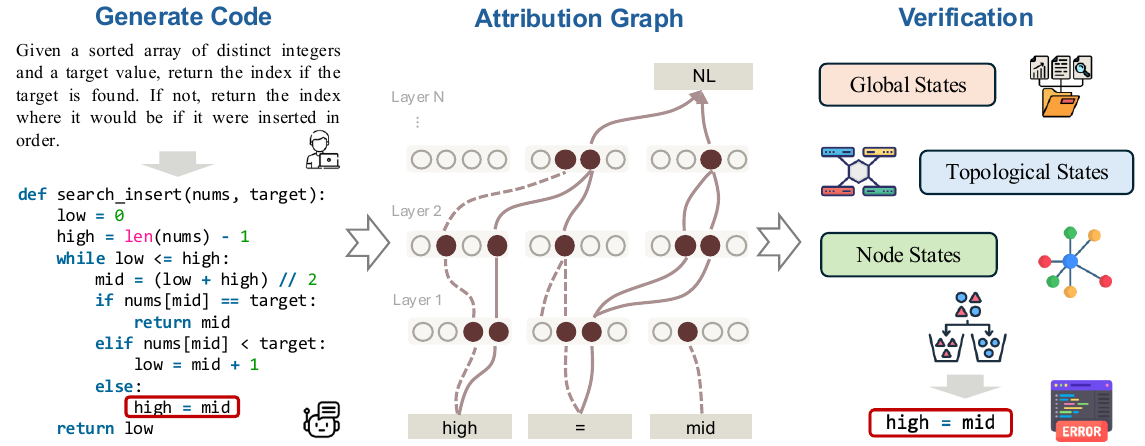}
  \vspace{-15pt}
  \caption{Overview of the \method{} framework. \method{} maps an LLM's internal dynamics into a line-level Attribution Graph to detect errors. By extracting structural features, including global, topological, and node states, the framework identifies latent structural fingerprints of validity. NL represents the next line.}
  \vspace{-5pt}
  \label{framework}
\end{figure*}

The fundamental premise of our framework \method{} (Figure~\ref{framework}) is that the correctness of a generated code sequence is intrinsically encoded within the \textit{latent algorithmic trajectory} of the language model. Traditional verification paradigms treat the model as an opaque function, focusing on extrinsic execution outcomes that often fail to capture the underlying causal mechanisms of logical fallacies. In contrast, building upon the mechanistic verification paradigm established by \citet{zhao2025verifying}, we propose a white-box framework that audits the structural integrity of the model's internal reasoning pathways. By projecting high-dimensional neural activations onto an interpretable basis, we extend their approach to identify the specific structural fingerprints of algorithmic success or failure in code, shifting the diagnostic focus from output observation to internal process auditing.

\subsection{Latent Algorithmic Tracing via Attribution Graphs}
To expose the hidden computational logic of code generation, we construct a step-level attribution graph $\mathcal{G}_i = (V_i, E_i)$ for each discrete logical step $i$. This graph serves as a formal causal map of the model's mechanistic reasoning pathway. As established in the preliminaries, we utilize a local replacement model where standard MLP layers are substituted by PLTs. This transformation projects the residual stream into a sparse, interpretable feature space $f^{(l)}$, where individual dimensions correspond to discrete algorithmic primitives. The resulting graph $\mathcal{G}_i$ captures the linear flow of information from input embeddings $V_{emb}$ through these latent features $V_{feat}$ to the final logit output. To account for the portion of the computation not captured by the interpretable features, we explicitly include error nodes $V_{err}$, defined as the residual difference between the true MLP output and its local PLT reconstruction:
\begin{equation}
    e^{(l)} = m^{(l)} - (W_{dec}^{(l)} f^{(l)} + b_{dec}^{(l)}).
\end{equation}
Including $V_{err}$ allows for a quantitative assessment of the mechanical faithfulness of the reasoning pathway, where the total influence of error nodes serves as a proxy for the epistemic uncertainty inherent in the model's latent algorithm.

\subsection{Structural Feature Extraction and Pathological Analysis}
Inspired by \citet{zhao2025verifying}, we hypothesize that logical errors in code generation manifest as structural anomalies within the attribution graph's topology. To capture these, we define a comprehensive feature extraction mapping $\Phi: \mathcal{G}_i \to \mathbf{x}_i$ that distills the complex topology of $\mathcal{G}_i$ into a fixed-size diagnostic vector $\mathbf{x}_i$. A complete enumeration of the statistics-, activation-, and topology-based features included in $\mathbf{x}_i$ is provided in Appendix~\ref{app:detailed_features}. This mapping systematically analyzes the circuit across multiple hierarchical dimensions of the computational trajectory.

The first dimension concerns the mechanical composition and faithfulness of the circuit. Beyond simple cardinality measures of the node set $V_i$ and edge set $E_i$, we define the \textit{error-to-feature influence ratio} $\eta_i$ to quantify the model's reliance on unexplained components. Let $w_{uv}$ denote the attribution weight of the edge from node $u$ to node $v$; the cumulative error influence is formulated as:
\begin{equation}
    \eta_i = \frac{\sum_{u \in V_{err}} \sum_{v \in V_i} |w_{uv}|}{\sum_{u \in V_{feat}} \sum_{v \in V_i} |w_{uv}|}.
\end{equation}
A high value of $\eta_i$ indicates a computation driven by opaque heuristics rather than interpretable algorithmic steps, suggesting a higher probability of logical fragility.

Furthermore, we analyze the global structure of the reasoning pathway. We compute the graph density $\rho_i = |E_i| / (|V_i|(|V_i|-1))$ and the number of connected components to evaluate whether the latent algorithm is executing a unified or fragmented process. To capture the hierarchical and modular nature of the computation, we calculate the average clustering coefficient $C_i$, which measures the degree to which nodes in the graph tend to cluster together:
\begin{equation}
C_i = \frac{1}{|V_i|} \sum_{v \in V_i}
\frac{
2 \left|
\left\{
\substack{
(u, w) \in E_i \\
u, w \in N(v) \\
(u, w) \in E_i
}
\right\}
\right|
}{
k_v (k_v - 1)
},
\end{equation}

where $N(v)$ is the neighborhood of node $v$ and $k_v$ is its degree. In the context of code generation, high modularity often corresponds to the execution of independent sub-tasks, while the presence of disjoint components may indicate fragmented reasoning where the model fails to integrate necessary context into the program state.

To identify critical information bottlenecks and logical hubs within the trajectory, we analyze the distribution of betweenness centrality $C_B(v)$ and degree centrality. For a node $v \in V_i$, the betweenness centrality represents the fraction of all shortest paths between node pairs that pass through $v$, formulated as:
\begin{equation}
    C_B(v) = \sum_{s \neq v \neq t \in V_i} \frac{\sigma_{st}(v)}{\sigma_{st}},
\end{equation}
where $\sigma_{st}$ is the total number of shortest paths from $s$ to $t$ and $\sigma_{st}(v)$ is the number of those paths passing through $v$. We aggregate these values into statistical moments, including the mean $\mu(C_B)$, standard deviation $\sigma(C_B)$, and maximum value $\max(C_B)$, to characterize the concentration of information flow. A robust algorithmic execution typically exhibits stable, high-centrality features that coordinate state-tracking, whereas an error-prone step may manifest as a fragmented bottleneck with no clear coordination hub.

Finally, we characterize attribution dynamics using a \textit{total logit attribution score}. Instead of relying on shallow connectivity, we aggregate the gradient-based attribution along all paths from a feature node to the final output logits. We extract the statistical profile of these total influence scores $S$, where $S_v$ represents the integrated gradient of the target probability with respect to feature $v$. This allows us to distinguish between features that merely activate and those that functionally drive the deep algorithmic processing essential for correct code generation.

\subsection{Diagnostic Classification and Structural Bug Diagnosis}
The final stage of our framework involves the training of a diagnostic classifier $h_\phi$ that maps the structural vector $\mathbf{x}_i$ to a correctness probability $\hat{y}_i = h_\phi(\mathbf{x}_i)$. We utilize a Gradient Boosting Decision Tree (GBDT;~\citealp{ke2017lightgbm}) architecture, which is particularly effective for handling the heterogeneous, tabular nature of the extracted topological features. The objective is to minimize the cross-entropy loss over a labeled dataset of correct and incorrect code generations:
\begin{equation}
    \mathcal{L}(\phi) = - \sum_{i} \left[ y_i \log(\hat{y}_i) + (1 - y_i) \log(1 - \hat{y}_i) \right].
\end{equation}
Beyond simple binary classification, the GBDT model provides feature importance rankings that enable \textit{structural bug diagnosis}. By identifying which topological properties are most predictive of failure, we can pinpoint the mechanical cause of a logical error. For instance, a high importance score for the error-to-feature ratio $\eta_i$ combined with low betweenness centrality for state-tracking features suggests that the model's failure stems from a lack of interpretable state maintenance. This transforms code verification from a post-hoc execution test into a proactive, mechanistic audit of the model's latent computational integrity.

\section{Experiments}

In this section, we empirically validate our central hypothesis: that \textit{code correctness is intrinsic to the model's latent computational topology and can be diagnosed without external execution.} We structure our evaluation around five primary research questions:

\begin{itemize}[leftmargin=*, itemsep=0pt, topsep=2pt]
    \item \textbf{RQ1 (Effectiveness):} Can \method{} accurately distinguish between correct and incorrect code across diverse programming languages?
    \item \textbf{RQ2 (Generalization):} Do the identified discriminative signals generalize across different programming syntax?
    \item \textbf{RQ3 (Scalability):}Does predictive power of internal signals remain robust as algorithmic complexity increases?
    \item \textbf{RQ4 (Visualization):} Are there visually distinct topological patterns in the internal circuits of correct versus incorrect code?
    \item \textbf{RQ5 (Causal Intervention):} Can targeted interventions on internal failure modes rectify errors in code generation?
\end{itemize}

\subsection{Evaluation Setup}
\label{expe}
\textbf{Datasets and Problem Domains.} To ensure cross-lingual robustness, we evaluate \method{} across Python, Java, and C++. We utilize the MBPP dataset~\cite{austin2021mbpp} as the benchmark for Python and curate semantically equivalent task sets for Java and C++ by translating MBPP specifications and generating solutions via Gemma-2-2B-it~\cite{team2024gemma2}. We use transcoder trained by~\cite{lieberum-etal-2024-gemma}. This alignment ensures that the model operates on identical algorithmic logic across disparate syntaxes. Ground truth labels are derived from GPT-4o~\cite{openai2024gpt4ocard}, where a step is defined as a correct code line. We emphasize that external supervision is used only to obtain evaluation labels and is not required by CodeCircuit during inference. Any functional or logical deviation within a line is treated as a positive instance of failure. Detailed dataset statistics and label distributions are provided in Appendix~\ref{app:dataset_stat}. 
The exact prompt templates used for code generation and line-wise labeling, as well as the attribution-graph extraction configuration, are shown in Appendix~\ref{app:prompt_templates} and~\ref{app:hyperparam}.

\textbf{Baselines.} We compare \method{} against a suite of established black-box and gray-box metrics. These include Temperature Scaling~\cite{ts}, Maximum Softmax Probability (MaxProb), Perplexity (PPL), Energy~\cite{liu2020energy}, and Entropy derived from final token logits as measures of surface-level confidence. We also include Chain-of-Embedding (CoE), such as real-space CoE-R, complex-space CoE-C variants~\cite{coe} and CoT-Kinetics~\cite{bi2025cot}, as a competitive gray-box baseline to evaluate the consistency of the reasoning process throughout the generation trajectory.

\textbf{Evaluation Metrics.} Following~\citet{zhao2025verifying}, we evaluate verifier performance using AUROC, FPR@95, and AUPR. Since the task focuses on identifying coding errors, incorrect steps are treated as the positive class throughout. Taken together, these metrics characterize ranking quality, error detection strength, and false alarm behavior.

\begin{table*}[t!]
\centering
\caption{Performance comparison of methods across different programming languages and baselines.}
\label{tab:cross_language}
\resizebox{\textwidth}{!}{
\begin{tabular}{llccccccccc}
\toprule
\multirow{2}{*}{\textbf{Paradigm}} & \multirow{2}{*}{\textbf{Method}} & \multicolumn{3}{c}{\textbf{Python}} & \multicolumn{3}{c}{\textbf{Java}} & \multicolumn{3}{c}{\textbf{C++}} \\
\cmidrule(lr){3-5} \cmidrule(lr){6-8} \cmidrule(lr){9-11}
& & AUROC $\uparrow$ & AUPR $\uparrow$ & FPR@95 $\downarrow$ & AUROC $\uparrow$ & AUPR $\uparrow$ & FPR@95 $\downarrow$ & AUROC $\uparrow$ & AUPR $\uparrow$ & FPR@95 $\downarrow$ \\
\midrule
\multirow{5}{*}{\textit{Black-Box}} 
& MaxProb        & 51.25 & 29.08 & 97.20 & 56.29 & 46.00 & 91.20 & 54.06 & 27.95 & 93.39 \\
& PPL            & 51.24 & 29.08 & 97.20 & 56.29 & 46.00 & 91.20 & 54.06 & 27.94 & 93.39 \\
& Entropy        & 51.42 & 29.04 & 96.88 & 56.56 & 46.34 & 89.96 & 53.16 & 27.98 & 93.39 \\
& Temp. Scaling  & 50.86 & 28.85 & 97.20 & 55.65 & 45.44 & 92.08 & 55.79 & 28.44 & 91.49 \\
& Energy         & 49.45 & 25.63 & 89.10 & 57.52 & 45.38 & 88.38 & 42.78 & 23.42 & 97.71 \\
\midrule
\multirow{4}{*}{\textit{Gray-Box}} 
& CoE-R          & 50.75 & 25.22 & 93.52 & 49.68 & 40.07 & 94.46 & 59.07 & 40.17 & 92.49 \\
& CoE-C          & 51.10 & 42.91 & 93.52 & 52.46 & 44.76 & 94.64 & 56.40 & 37.72 & 92.61 \\
& CoT-Kinetics   & 51.41 & 29.04 & 96.88 & 56.57 & 46.34 & 89.96 & 53.16 & 27.98 & 93.39 \\
\midrule
\textbf{\textit{White-Box}} & \textbf{Ours} & \textbf{79.89} & \textbf{54.77} & \textbf{77.84} & \textbf{68.06} & \textbf{56.83} & \textbf{79.66} & \textbf{72.21} & \textbf{64.55} & \textbf{80.51} \\
\bottomrule
\end{tabular}
}
\end{table*}

\begin{table}[h]
\centering
\caption{Performance comparison of different methods across Python, C++, and Java test sets. Results are evaluated using AUROC, AUPR, and FPR@95 metrics.}
\label{tab:cross_domain}
\resizebox{0.48\textwidth}{!}{
\begin{tabular}{llccc}
\toprule
\textbf{Test Set} & \textbf{Method (Train Set)} & \multicolumn{3}{c}{\textbf{Metrics}} \\
\cmidrule(lr){3-5}
& & \textbf{AUROC $\uparrow$} & \textbf{AUPR $\uparrow$} & \textbf{FPR@95 $\downarrow$} \\
\midrule
\multirow{4}{*}{\textbf{Python}} 
& Baseline (MaxProb) & 51.25 & 29.08 & 97.20\\
& \method{} (Java) & 55.42 & 34.15 & 94.10 \\
& \method{} (C++)    & 53.18 & 31.50 & 95.82 \\
\cmidrule(lr){2-5}
& \emph{\method{} (Python)} & \emph{79.89} & \emph{54.77} & \emph{77.84} \\
\midrule
\multirow{4}{*}{\textbf{Java}} 
& Baseline (MaxProb) & 56.29 & 46.00 & 91.20\\
& \method{} (Python) & 62.15 & 48.33 & 88.45 \\
& \method{} (C++)    & 58.74 & 47.12 & 90.15 \\
\cmidrule(lr){2-5}
& \emph{\method{} (Java)} & \emph{68.06} & \emph{56.83} & \emph{79.66}\\
\midrule
\multirow{4}{*}{\textbf{C++}} 
& Baseline (MaxProb) & 54.06 & 27.95 & 93.39\\
& \method{} (Java)   & 59.33 & 36.80 & 88.24 \\
& \method{} (Python) & 57.10 & 32.55 & 91.12 \\
\cmidrule(lr){2-5}
& \emph{\method{} (C++)} & \emph{72.21} & \emph{64.55} & \emph{80.51} \\
\bottomrule
\end{tabular}}
\end{table}

\subsection{Main Performance}

\textbf{Effectiveness.} 
We first evaluate the fundamental efficacy of \method{} in identifying code errors. This analysis addresses whether internal circuit pathologies provide a more reliable signal for correctness than surface-level heuristics (RQ1). As shown in Table~\ref{tab:cross_language}, \method{} outperforms all baseline paradigms across Python, Java, and C++, demonstrating the superior reliability of internal circuit pathologies over surface-level heuristics. Specifically, our approach achieves an AUROC of 79.89 in Python, representing a substantial improvement over the best-performing black-box and gray-box methods, which hover around 51.42. This trend is consistent across other metrics and languages; for instance, we achieve an AUPR of 54.77 in Python and 56.83 in Java, nearly doubling the performance of traditional methods like MaxProb or CoT-Kinetics in many cases. Furthermore, our method markedly reduces the FPR@95, maintaining values as low as 80.51 in C++, whereas baselines frequently exceed 90. These results highlight that by leveraging internal model states, we can distinguish correct from erroneous code with far greater precision than methods relying on output probabilities or chain-of-thought consistency.

\textbf{Generalization.} We investigate the cross-programming language invariance of these structural signatures by conducting transfer experiments across Python, Java, and C++ (RQ2). As shown in Table \ref{tab:cross_domain}, our method exhibits robust zero-shot generalization capabilities, consistently outperforming the surface-level baseline (MaxProb) even when the probe is trained on a different programming language. For instance, when testing on the Java set, the probe trained on Python achieves an AUROC of 62.15, marking a 5.86\% absolute improvement over the Java-specific MaxProb baseline (56.29). We observe a nuanced transferability across the three languages; notably, the Java-trained probe generalizes effectively to both C++ (59.33 AUROC) and Python (55.42 AUROC), suggesting that the structural patterns captured from Java may encapsulate more fundamental logic common to imperative languages. While a performance gap naturally exists between cross-domain transfer and in-domain training, the consistent superiority of our method over the baseline confirms that code errors manifest through a set of universal structural pathologies. These results suggest that our white-box approach captures fundamental reasoning inconsistencies that transcend language-specific syntax.

\begin{figure*}[t]
  \centering
  \includegraphics[width=\textwidth]{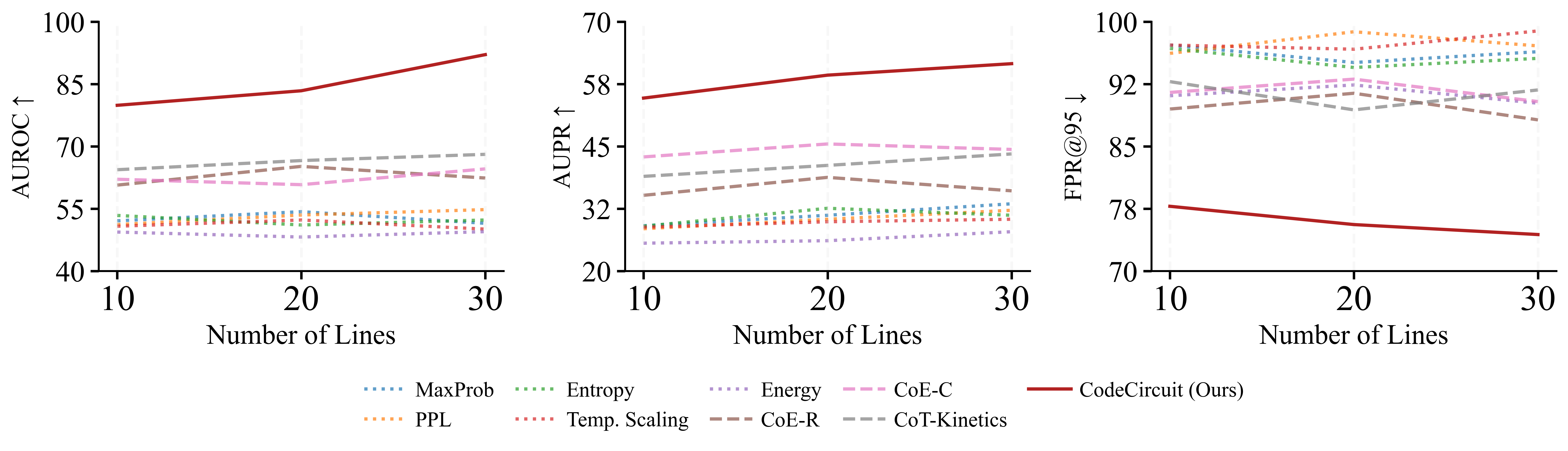}
  \vspace{-20pt}
\caption{Predictor performance on Python tasks by difficulty, showing~\method{}’s advantage as complexity increases.}
  \vspace{-5pt}
  \label{scal}
\end{figure*}

\begin{figure*}[t]
  \centering
  \includegraphics[width=0.9\textwidth]{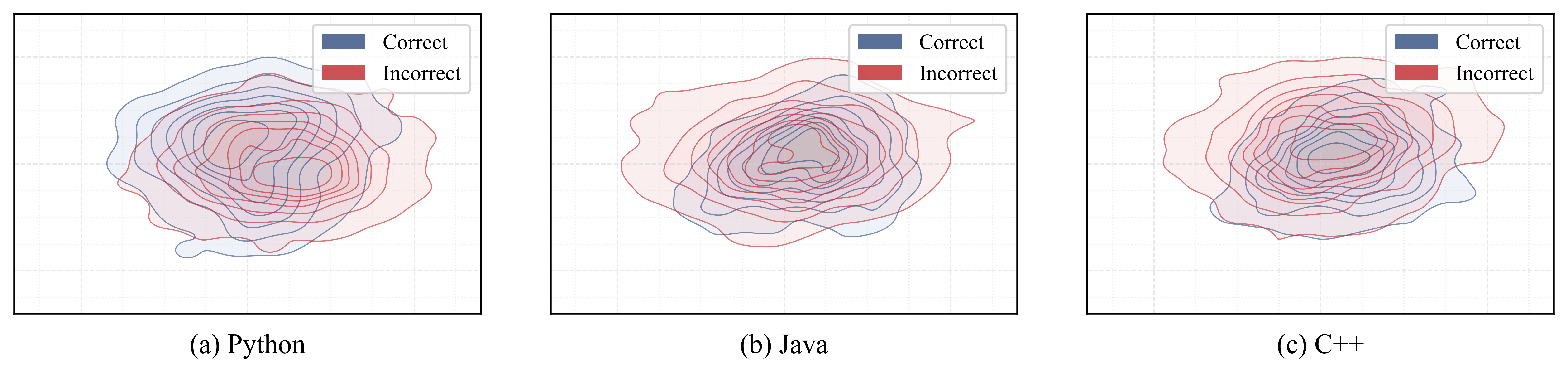}
  \vspace{-5pt}
  \caption{Feature distributions after PCA for correct (blue) and incorrect (red) code generation in different programming languages.}
  \vspace{-5pt}
  \label{vis}
\end{figure*}

\begin{table*}[h]
\centering
\small
\renewcommand{\arraystretch}{1.2} 
\begin{tabularx}{\textwidth}{XX}
\toprule
\textbf{Before Intervention (Incorrect)} & \textbf{After Intervention (Correct)} \\ \midrule
\texttt{def search\_insert(nums, target):} & \texttt{def search\_insert(nums, target):} \\
\texttt{\hspace{2em}low, high = 0, len(nums) - 1} & \texttt{\hspace{2em}low, high = 0, len(nums) - 1} \\
\texttt{\hspace{2em}while low <= high:} & \texttt{\hspace{2em}while low <= high:} \\
\texttt{\hspace{4em}mid = (low + high) // 2} & \texttt{\hspace{4em}mid = (low + high) // 2} \\
\texttt{\hspace{4em}if nums[mid] == target:} & \texttt{\hspace{4em}if nums[mid] == target:} \\
\texttt{\hspace{6em}return mid} & \texttt{\hspace{6em}return mid} \\
\texttt{\hspace{4em}elif nums[mid] < target:} & \texttt{\hspace{4em}elif nums[mid] < target:} \\
\texttt{\hspace{6em}low = mid + 1} & \texttt{\hspace{6em}low = mid + 1} \\
\texttt{\hspace{4em}else:} & \texttt{\hspace{4em}else:} \\
\texttt{\hspace{6em}}\colorbox{highlightRed}{\texttt{high = mid}} & \texttt{\hspace{6em}}\colorbox{highlightBlue}{\texttt{high = mid - 1}} \\
\texttt{\hspace{2em}return low} & \texttt{\hspace{2em}return low} \\
\midrule
\textit{Trace:} The model fails to decrement \texttt{high}, leading to a potential infinite loop or incorrect insertion index. & \textit{Trace:} Suppressing the ``greedy'' feature allows the model to correctly update the boundary. \\
\bottomrule
\end{tabularx}
\caption{Side-by-side comparison of a code reasoning trace \colorbox{highlightRed}{before} and \colorbox{highlightBlue}{after} causal intervention. The highlight indicates the point of divergence where suppressing a single \textit{greedy-matching} transcoder feature corrects the model's internal computational path and restores logical validity.}
\label{tab:code_intervention}
\end{table*}

\textbf{Scalability}. 
To test the robustness of these signals, we analyze how predictive power scales as code complexity and physical length (number of lines $N \in \{10, 20, 30\}$) increase (RQ3). As shown in the Figure~\ref{scal}, the performance advantage of \method{} becomes significantly more pronounced as task difficulty grows, with AUROC rising from approximately 80 to nearly 92. While surface-level baselines like MaxProb remain stagnant, our method capitalizes on the richer internal activations provided by longer sequences to more accurately identify circuit pathologies. This suggests that unlike gray-box methods that struggle with consistency, our white-box approach becomes increasingly reliable in more demanding and complex computational contexts.

\subsection{Visual Analysis (RQ4)}
To move beyond binary prediction, we conduct a qualitative and fine-grained analysis of the attribution graphs to uncover the ``structural fingerprints'' of specific failure modes. Following~\citet{zhao2025verifying}, we visualize the separation between correct and erroneous computational trajectories using manifold projection techniques and case studies of individual circuits in Figure~\ref{vis}. The plots reveal that correct steps often form a dense subset within the broader distribution of incorrect steps. This spatial overlap suggests many failures are structural near misses, maintaining similar graph topologies to valid code. Nevertheless, a distinct zone of computational integrity exists where only correct samples reside. This observation is consistent with the findings of~\citet{zhao2025verifying} for reasoning tasks. This consistency across languages confirms that our graph features capture universal signatures of correctness that transcend specific programming language.


\subsection{Causal Intervention (RQ5)}

Finally, we move beyond passive verification to test the causal nature of the identified structural features. Consistent with recent findings by \citet{zhao2025verifying} in the domain of reasoning, we hypothesize that the anomalies detected by \method{} are not merely correlational byproducts but active drivers of the model's generation. To validate this, we perform a ``mechanistic patching'' experiment—intervening on specific nodes in the attribution graph to correct algorithmic logic in real-time.

We illustrate this via a case study on a classic boundary condition failure. Given a prompt to implement a binary search algorithm, the model generated the incorrect update logic \texttt{high = mid}, a common bug that leads to infinite loops, instead of the correct \texttt{high = mid - 1} (Table \ref{tab:code_intervention}). \method{} flagged this step as structurally anomalous. Investigating the attribution graph revealed that the error was driven by a specific transcoder feature in the middle layers, which semantically corresponded to a ``greedy matching'' heuristic that overrode the precise arithmetic state maintenance required for the loop invariant.

To test causality, we performed a targeted intervention using activation clamping. We re-ran the generation and verified that by placing a forward hook to suppress (zero out) this maladaptive feature, the model's trajectory was immediately corrected. As shown in Table~\ref{tab:code_intervention}, the intervention successfully steered the model to generate \texttt{high = mid - 1} and proceed to a functionally correct solution. We also observed successful corrections via feature \textit{amplification}, specifically by boosting under-active features associated with variable binding context.

These results corroborate the causal link between latent circuit activation and reasoning correctness established by \citet{zhao2025verifying}, while extending the phenomenon to the strict syntactic and logical constraints of programming. This serves as a proof-of-concept for \textit{latent debugging}: the ability to fix software bugs not by editing the source code, but by modulating the neural circuitry that generates it.








\section{Related Work}
\subsection{Mechanistic Interpretability}
Mechanistic interpretability~\cite{openai_mc,mc1,rai2024practical} aims to reverse-engineer neural networks into their underlying computational algorithms. The field was largely formalized by the ``circuits" framework \cite{olah2020zoom}, which posits that features are the fundamental units of representation and are connected by weights to form circuits. In the context of Transformers, \citet{elhage2021mathematical} introduced a mathematical framework that identified ``induction heads'' as the mechanism for in-context learning. This paradigm has enabled the discovery of specific circuits for tasks such as indirect object identification \cite{wang2022interpretability} and modular arithmetic \cite{nanda2023progress}. To address the challenge of polysemanticity, where single neurons represent multiple unrelated concepts \cite{olah2017feature}, recent research has leveraged the Linear Representation Hypothesis \cite{elhage2022softmax} and Sparse Autoencoders to decompose activations into monosemantic features \cite{bricken2023towards, cunningham2023sparse}. Furthermore, to bridge the gap between manual inspection and large-scale models, automated discovery methods such as ACDC~\cite{conmy2023towards} and LLM-based explanations \cite{bills2023language} have been proposed.

\subsection{LLM-based Code Generation}
LLMs have rapidly advanced from basic code completion utilities to sophisticated systems for automated software engineering~\cite{liu2024deepseek, he2025visplay, li2023starcoder, wang2023codet5plus, hui2024qwen25coder}. The initial phase of this evolution was marked by the development of foundational models, such as CodeLlama \cite{Roziere2023CodeLlama} and DeepSeek-Coder \cite{Guo2024DeepSeekCoder}, which established new performance baselines. These models were typically trained on massive, high-quality code corpora and leveraged long-context windows to capture extensive code dependencies.
A subsequent research direction has focused on enhancing the functional correctness of generated code, moving beyond conventional open-loop generation. This involves the integration of execution-based feedback and self-debugging mechanisms \cite{Chen2023SelfDebug, Gao2024SelfPlanning}, where the model iteratively refines its proposed solution based on signals derived from compilers, linters, or test case execution.
More recently, the scope of LLM-based code generation has expanded from isolated code snippets to full project-level development. The emergence of agentic frameworks, capable of autonomously addressing real-world software issues, has been a significant development. The capabilities of these agents are rigorously evaluated using benchmarks like SWE-bench \cite{Jimenez2023SWEBench} and platforms such as OpenDevin \cite{Wang2024OpenDevin}.

\section{Conclusion}
This work asks a single question: whether the correctness of LLM-generated code can be inferred from the model’s internal computation, without relying on execution, tests, or external judges. This question is motivated by the growing use of LLMs in settings where external evaluation is costly or incomplete, and where reliability must be assessed from the generation process itself rather than from downstream outcomes.
Through a systematic analysis based on line-level attribution graphs, we show that this is indeed possible. Correct and incorrect code exhibit consistent and distinguishable internal structures across programming languages, indicating that code correctness is encoded in the model’s internal representations. These results suggest that internal analysis provides a viable and general approach to assessing code correctness, and open the door to reliability evaluation methods grounded directly in model behavior rather than external supervision.

\section*{Impact Statement}

This work contributes to the advancement of reliable and interpretable machine learning, specifically in the domain of automated software engineering. As LLMs are increasingly deployed to generate code for critical infrastructure and applications, ensuring their correctness is of paramount societal importance. Our proposed framework, \method, advances this goal by shifting verification from opaque, black-box outcomes to transparent, mechanistic processes. By exposing the internal ``circuits'' of code generation, this work aids in the early detection of logical failures and software vulnerabilities that standard execution-based testing might miss. Furthermore, our focus on interpretability aligns with the broader goal of developing trustworthy AI systems that are understandable by humans. However, we note that while mechanistic verification improves reliability, it should currently serve as an aid to, rather than a replacement for, human review and rigorous software testing protocols.


\bibliography{example_paper}

@article{anthropic_attributiongraph,
  author={Ameisen, Emmanuel and Lindsey, Jack and Pearce, Adam and Gurnee, Wes and Turner, Nicholas L. and Chen, Brian and Citro, Craig and Abrahams, David and Carter, Shan and Hosmer, Basil and Marcus, Jonathan and Sklar, Michael and Templeton, Adly and Bricken, Trenton and McDougall, Callum and Cunningham, Hoagy and Henighan, Thomas and Jermyn, Adam and Jones, Andy and Persic, Andrew and Qi, Zhenyi and Ben Thompson, T. and Zimmerman, Sam and Rivoire, Kelley and Conerly, Thomas and Olah, Chris and Batson, Joshua},
  title={Circuit Tracing: Revealing Computational Graphs in Language Models},
  journal={Transformer Circuits Thread},
  year={2025},
  url={https://transformer-circuits.pub/2025/attribution-graphs/methods.html}
}

@inproceedings{dunefsky2024transcoders,
title={Transcoders find interpretable {LLM} feature circuits},
author={Jacob Dunefsky and Philippe Chlenski and Neel Nanda},
booktitle={The Thirty-eighth Annual Conference on Neural Information Processing Systems},
year={2024},
url={https://openreview.net/forum?id=J6zHcScAo0}
}

@article{zhao2025verifying,
  title={Verifying Chain-of-Thought Reasoning via Its Computational Graph},
  author={Zhao, Zheng and Koishekenov, Yeskendir and Yang, Xianjun and Murray, Naila and Cancedda, Nicola},
  journal={arXiv preprint arXiv:2510.09312},
  year={2025}
}

@inproceedings{lieberum-etal-2024-gemma,
    title = "Gemma Scope: Open Sparse Autoencoders Everywhere All At Once on Gemma 2",
    author = "Lieberum, Tom  and
      Rajamanoharan, Senthooran  and
      Conmy, Arthur  and
      Smith, Lewis  and
      Sonnerat, Nicolas  and
      Varma, Vikrant  and
      Kramar, Janos  and
      Dragan, Anca  and
      Shah, Rohin  and
      Nanda, Neel",
    editor = "Belinkov, Yonatan  and
      Kim, Najoung  and
      Jumelet, Jaap  and
      Mohebbi, Hosein  and
      Mueller, Aaron  and
      Chen, Hanjie",
    booktitle = "Proceedings of the 7th BlackboxNLP Workshop: Analyzing and Interpreting Neural Networks for NLP",
    month = nov,
    year = "2024",
    address = "Miami, Florida, US",
    publisher = "Association for Computational Linguistics",
    url = "https://aclanthology.org/2024.blackboxnlp-1.19/",
    doi = "10.18653/v1/2024.blackboxnlp-1.19",
    pages = "278--300"
}

@misc{openai2024gpt4ocard,
      title={GPT-4o System Card}, 
      author={OpenAI and others},
      year={2024},
      eprint={2410.21276},
      archivePrefix={arXiv},
      primaryClass={cs.CL},
      url={https://arxiv.org/abs/2410.21276}, 
}

@misc{circuit-tracer,
  author = {Hanna, Michael and Piotrowski, Mateusz and Lindsey, Jack and Ameisen, Emmanuel},
  title = {circuit-tracer},
  howpublished = {\url{https://github.com/safety-research/circuit-tracer}},
  note = {The first two authors contributed equally and are listed alphabetically.},
  year = {2025}
}

@article{he2025visplay,
  title={VisPlay: Self-Evolving Vision-Language Models from Images},
  author={He, Yicheng and Huang, Chengsong and Li, Zongxia and Huang, Jiaxin and Yang, Yonghui},
  journal={arXiv preprint arXiv:2511.15661},
  year={2025}
}

@article{liu2024deepseek,
  title={Deepseek-v3 technical report},
  author={Liu, Aixin and Feng, Bei and Xue, Bing and Wang, Bingxuan and Wu, Bochao and Lu, Chengda and Zhao, Chenggang and Deng, Chengqi and Zhang, Chenyu and Ruan, Chong and others},
  journal={arXiv preprint arXiv:2412.19437},
  year={2024}
}

@article{liu2023your,
  title={Is your code generated by chatgpt really correct? rigorous evaluation of large language models for code generation},
  author={Liu, Jiawei and Xia, Chunqiu Steven and Wang, Yuyao and Zhang, Lingming},
  journal={Advances in Neural Information Processing Systems},
  volume={36},
  pages={21558--21572},
  year={2023}
}

@article{liu2024evaluating,
  title={Evaluating language models for efficient code generation},
  author={Liu, Jiawei and Xie, Songrun and Wang, Junhao and Wei, Yuxiang and Ding, Yifeng and Zhang, Lingming},
  journal={arXiv preprint arXiv:2408.06450},
  year={2024}
}

@inproceedings{khan2024xcodeeval,
  title={Xcodeeval: An execution-based large scale multilingual multitask benchmark for code understanding, generation, translation and retrieval},
  author={Khan, Mohammad Abdullah Matin and Bari, M Saiful and Do, Xuan Long and Wang, Weishi and Parvez, Md Rizwan and Joty, Shafiq},
  booktitle={Proceedings of the 62nd Annual Meeting of the Association for Computational Linguistics (Volume 1: Long Papers)},
  pages={6766--6805},
  year={2024}
}

@article{dong2025codescore,
  title={Codescore: Evaluating code generation by learning code execution},
  author={Dong, Yihong and Ding, Jiazheng and Jiang, Xue and Li, Ge and Li, Zhuo and Jin, Zhi},
  journal={ACM Transactions on Software Engineering and Methodology},
  volume={34},
  number={3},
  pages={1--22},
  year={2025},
  publisher={ACM New York, NY}
}

@inproceedings{wang2023execution,
  title={Execution-based evaluation for open-domain code generation},
  author={Wang, Zhiruo and Zhou, Shuyan and Fried, Daniel and Neubig, Graham},
  booktitle={Findings of the Association for Computational Linguistics: EMNLP 2023},
  pages={1271--1290},
  year={2023}
}

@article{bui2025correctness,
  title={Correctness Assessment of Code Generated by Large Language Models Using Internal Representations},
  author={Bui, Tuan-Dung and Vu, Thanh Trong and Nguyen, Thu-Trang and Nguyen, Son and Vo, Hieu Dinh},
  journal={arXiv preprint arXiv:2501.12934},
  year={2025}
}

@inproceedings{huang2025risk,
  title={Risk assessment framework for code llms via leveraging internal states},
  author={Huang, Yuheng and Ma, Lei and Nishikino, Keizaburo and Akazaki, Takumi},
  booktitle={Proceedings of the 33rd ACM International Conference on the Foundations of Software Engineering},
  pages={432--443},
  year={2025}
}

@inproceedings{patel2025planning,
  title        = {Planning a Large Language Model for Static Detection of Runtime Errors in Code Snippets},
  author       = {Patel, Smit and Yadavally, Aashish and Dhulipala, Hridya and Nguyen, Tien},
  booktitle    = {Proceedings of the IEEE/ACM 47th International Conference on Software Engineering (ICSE 2025)},
  pages        = {872--884},
  year         = {2025},
  doi          = {10.1109/ICSE55347.2025.00102}
}

@article{mc1,
  title={Opening the ai black box: program synthesis via mechanistic interpretability},
  author={Michaud, Eric J and Liao, Isaac and Lad, Vedang and Liu, Ziming and Mudide, Anish and Loughridge, Chloe and Guo, Zifan Carl and Kheirkhah, Tara Rezaei and Vukeli{\'c}, Mateja and Tegmark, Max},
  journal={arXiv preprint arXiv:2402.05110},
  year={2024}
}

@article{mc2,
  title={Open problems in mechanistic interpretability},
  author={Sharkey, Lee and Chughtai, Bilal and Batson, Joshua and Lindsey, Jack and Wu, Jeff and Bushnaq, Lucius and Goldowsky-Dill, Nicholas and Heimersheim, Stefan and Ortega, Alejandro and Bloom, Joseph and others},
  journal={arXiv preprint arXiv:2501.16496},
  year={2025}
}

@article{mc3,
  title={A practical review of mechanistic interpretability for transformer-based language models},
  author={Rai, Daking and Zhou, Yilun and Feng, Shi and Saparov, Abulhair and Yao, Ziyu},
  journal={arXiv preprint arXiv:2407.02646},
  year={2024}
}

@article{anthropic_mc,
   title={Scaling Monosemanticity: Extracting Interpretable Features from Claude 3 Sonnet},
   author={Templeton, Adly and Conerly, Tom and Marcus, Jonathan and Lindsey, Jack and Bricken, Trenton and Chen, Brian and Pearce, Adam and Citro, Craig and Ameisen, Emmanuel and Jones, Andy and Cunningham, Hoagy and Turner, Nicholas L and McDougall, Callum and MacDiarmid, Monte and Freeman, C. Daniel and Sumers, Theodore R. and Rees, Edward and Batson, Joshua and Jermyn, Adam and Carter, Shan and Olah, Chris and Henighan, Tom},
   year={2024},
   journal={Transformer Circuits Thread},
   url={https://transformer-circuits.pub/2024/scaling-monosemanticity/index.html}
}

@misc{openai_mc,
      title={Weight-sparse transformers have interpretable circuits}, 
      author={Leo Gao and Achyuta Rajaram and Jacob Coxon and Soham V. Govande and Bowen Baker and Dan Mossing},
      year={2025},
      eprint={2511.13653},
      archivePrefix={arXiv},
      url={https://arxiv.org/abs/2511.13653}, 
}

@article{olah2020zoom,
  title={Zoom in: An introduction to circuits},
  author={Olah, Chris and Cammarata, Nick and Schubert, Ludwig and Goh, Gabriel and Petrov, Michael and Carter, Shan},
  journal={Distill},
  volume={5},
  number={3},
  pages={e00024--001},
  year={2020}
}

@article{elhage2021mathematical,
  title={A mathematical framework for transformer circuits},
  author={Elhage, Nelson and Nanda, Neel and Olsson, Catherine and Henighan, Tom and Joseph, Nicholas and Mann, Ben and Askell, Amanda and Bai, Yuntao and Chen, Anna and Conerly, Tom and others},
  journal={Transformer Circuits Thread},
  volume={1},
  number={1},
  pages={12},
  year={2021}
}

@article{wang2022interpretability,
  title={Interpretability in the wild: a circuit for indirect object identification in gpt-2 small},
  author={Wang, Kevin and Variengien, Alexandre and Conmy, Arthur and Shlegeris, Buck and Steinhardt, Jacob},
  journal={arXiv preprint arXiv:2211.00593},
  year={2022}
}

@article{nanda2023progress,
  title={Progress measures for grokking via mechanistic interpretability},
  author={Nanda, Neel and Chan, Lawrence and Lieberum, Tom and Smith, Jess and Steinhardt, Jacob},
  journal={arXiv preprint arXiv:2301.05217},
  year={2023}
}

@misc{elhage2022softmax,
  title={Softmax linear units. transformer circuits thread},
  author={Elhage, N and Hume, T and Olsson, C and Nanda, N and Henighan, T and Johnston, S and ElShowk, S and Joseph, N and DasSarma, N and Mann, B and others},
  year={2022}
}

@article{bricken2023towards,
  title={Towards monosemanticity: Decomposing language models with dictionary learning},
  author={Bricken, Trenton and Templeton, Adly and Batson, Joshua and Chen, Brian and Jermyn, Adam and Conerly, Tom and Turner, Nick and Anil, Cem and Denison, Carson and Askell, Amanda and others},
  journal={Transformer Circuits Thread},
  volume={2},
  year={2023}
}

@article{conmy2023towards,
  title={Towards automated circuit discovery for mechanistic interpretability},
  author={Conmy, Arthur and Mavor-Parker, Augustine and Lynch, Aengus and Heimersheim, Stefan and Garriga-Alonso, Adri{\`a}},
  journal={Advances in Neural Information Processing Systems},
  volume={36},
  pages={16318--16352},
  year={2023}
}

@article{olah2017feature,
  title={Feature visualization},
  author={Olah, Chris and Mordvintsev, Alexander and Schubert, Ludwig},
  journal={Distill},
  volume={2},
  number={11},
  pages={e7},
  year={2017}
}

@article{cunningham2023sparse,
  title={Sparse autoencoders find highly interpretable features in language models},
  author={Cunningham, Hoagy and Ewart, Aidan and Riggs, Logan and Huben, Robert and Sharkey, Lee},
  journal={arXiv preprint arXiv:2309.08600},
  year={2023}
}

@misc{bills2023language,
 title={Language models can explain neurons in language models},
 author={
    Bills, Steven and Cammarata, Nick and Mossing, Dan and Tillman, Henk and Gao, Leo and Goh, Gabriel and Sutskever, Ilya and Leike, Jan and Wu, Jeff and Saunders, William
 },
 year={2023},
 howpublished = {\url{https://openaipublic.blob.core.windows.net/neuron-explainer/paper/index.html}}
}

@article{llm_asjudge_1,
  title={LLM-as-a-Judge for Software Engineering: Literature Review, Vision, and the Road Ahead},
  author={He, Junda and Shi, Jieke and Zhuo, Terry Yue and Treude, Christoph and Sun, Jiamou and Xing, Zhenchang and Du, Xiaoning and Lo, David},
  journal={arXiv preprint arXiv:2510.24367},
  year={2025}
}

@article{llm_asjudge_2,
  title={Codejudgebench: Benchmarking llm-as-a-judge for coding tasks},
  author={Jiang, Hongchao and Chen, Yiming and Cao, Yushi and Lee, Hung-yi and Tan, Robby T},
  journal={arXiv preprint arXiv:2507.10535},
  year={2025}
}

@inproceedings{ts,
  title={Long horizon temperature scaling},
  author={Shih, Andy and Sadigh, Dorsa and Ermon, Stefano},
  booktitle={International conference on machine learning},
  pages={31422--31434},
  year={2023},
  organization={PMLR}
}

@article{coe,
  title={Latent space chain-of-embedding enables output-free llm self-evaluation},
  author={Wang, Yiming and Zhang, Pei and Yang, Baosong and Wong, Derek F and Wang, Rui},
  journal={arXiv preprint arXiv:2410.13640},
  year={2024}
}

@article{liu2020energy,
  title={Energy-based out-of-distribution detection},
  author={Liu, Weitang and Wang, Xiaoyun and Owens, John and Li, Yixuan},
  journal={Advances in neural information processing systems},
  volume={33},
  pages={21464--21475},
  year={2020}
}

@article{bi2025cot,
  title={Cot-kinetics: A theoretical modeling assessing lrm reasoning process},
  author={Bi, Jinhe and Yan, Danqi and Wang, Yifan and Huang, Wenke and Chen, Haokun and Wan, Guancheng and Ye, Mang and Xiao, Xun and Schuetze, Hinrich and Tresp, Volker and others},
  journal={arXiv preprint arXiv:2505.13408},
  year={2025}
}

@article{Roziere2023CodeLlama,
  title={Code Llama: Open Foundation Models for Code},
  author={Rozi{\`e}re, Baptiste and Gehring, Jonas and Gloeckle, Fabian and Sootla, Sida and Gat, Itai and Tan, Xiao E and others},
  journal={arXiv preprint arXiv:2308.12950},
  year={2023}
}

@article{Guo2024DeepSeekCoder,
  title={DeepSeek-Coder: When the Large Language Model Meets Programming--The Rise of Code Intelligence},
  author={Guo, Daya and Zhu, Qihao and Yang, Deyi and Xie, Zhaofeng and Dong, Kun and Zhang, Wenshen and others},
  journal={arXiv preprint arXiv:2401.14196},
  year={2024}
}

@inproceedings{Chen2023SelfDebug,
  title={Teaching Large Language Models to Self-Debug},
  author={Chen, Xin and Lin, Maxwell and Sch{\"a}rli, Niki and Zhou, Denny},
  booktitle={International Conference on Learning Representations (ICLR)},
  year={2024}
}

@article{Gao2024SelfPlanning,
  title={Self-planning code generation with large language models},
  author       = {Jiang, Xue and Dong, Jiazheng and Wang, Yihong and Zheng, Yifan and Shang, Lei and Li, Ge and Jin, Zhi and Jiao, Li},
  journal      = {ACM Transactions on Software Engineering and Methodology},
  year         = {2024},
  volume       = {33},
  number       = {7},
  doi          = {10.1145/3672456}
}

@article{Jimenez2023SWEBench,
  title={SWE-bench: Can language models resolve real-world github issues?},
  author={Jimenez, Carlos E and Yang, John and Wettig, Alexander and Yao, Shunyu and Pei, Kexin and Niranjan, Ofir and others},
  journal={arXiv preprint arXiv:2310.06770},
  year={2023}
}

@article{Wang2024OpenDevin,
  title={OpenDevin: An Open Platform for AI Software Developers as Generalist Agents},
  author={Wang, Xingyao and Li, Boxuan and Neubig, Graham},
  journal={arXiv preprint arXiv:2407.16741},
  year={2024}
}

@inproceedings{zhang2025fsa_cot,
  title={Finite State Automata Inside Transformers with Chain-of-Thought: A Mechanistic Study on State Tracking},
  author={Zhang, Yifan and Du, Wenyu and Jin, Dongming and Fu, Jie and Jin, Zhi},
  booktitle={Proceedings of the 63rd Annual Meeting of the Association for Computational Linguistics (ACL)},
  year={2025}
}

@article{chen2021codex_humaneval,
  title={Evaluating Large Language Models Trained on Code},
  author={Chen, Mark and Tworek, Jerry and Jun, Heewoo and Yuan, Qiming and Pinto, Henrique Ponde de Oliveira and Kaplan, Jared and Edwards, Harri and Burda, Yuri and Joseph, Nicholas and Brockman, Greg and others},
  journal={arXiv preprint arXiv:2107.03374},
  year={2021}
}

@article{austin2021mbpp,
  title={Program Synthesis with Large Language Models},
  author={Austin, Jacob and Odena, Augustus and Nye, Maxwell and Bosma, Maarten and Michalewski, Henryk and Dohan, David and Jiang, Ellen and Cai, Carrie and Terry, Michael and Le, Quoc V. and others},
  journal={arXiv preprint arXiv:2108.07732},
  year={2021}
}

@inproceedings{hendrycks2021apps,
  title={Measuring Coding Challenge Competence With {APPS}},
  author={Hendrycks, Dan and Basart, Steven and Kadavath, Saurav and Mazeika, Mantas and Arora, Akul and Guo, Ethan and Burns, Collin and Puranik, Samir and He, Horace and Song, Dawn and others},
  booktitle={Advances in Neural Information Processing Systems (NeurIPS) Datasets and Benchmarks Track},
  year={2021}
}

@inproceedings{meng2022rome,
  title={Locating and Editing Factual Associations in {GPT}},
  author={Meng, Kevin and Bau, David and Andonian, Alex and Belinkov, Yonatan},
  booktitle={Advances in Neural Information Processing Systems (NeurIPS)},
  year={2022}
}

@inproceedings{meng2023memit,
  title={Mass-Editing Memory in a Transformer},
  author={Meng, Kevin and Bau, David and Andonian, Alex and Belinkov, Yonatan},
  booktitle={International Conference on Learning Representations (ICLR)},
  year={2023}
}

@inproceedings{madaan2023selfrefine,
  title={Self-Refine: Iterative Refinement with Self-Feedback},
  author={Madaan, Aman and Tandon, Niket and Gupta, Prakhar and Hallinan, Skyler and Gao, Luyu and Wiegreffe, Sarah and Alon, Uri and Dziri, Nouha and Prabhumoye, Shrimai and Yang, Shuo and others},
  booktitle={Advances in Neural Information Processing Systems (NeurIPS)},
  year={2023}
}

@inproceedings{shinn2023reflexion,
  title={Reflexion: Language Agents with Verbal Reinforcement Learning},
  author={Shinn, Noah and Cassano, Federico and Gopinath, Ashwin and Narasimhan, Karthik and Yao, Shunyu},
  booktitle={Advances in Neural Information Processing Systems (NeurIPS)},
  year={2023}
}

@article{rai2024practical,
  title={A Practical Review of Mechanistic Interpretability for Transformer-Based Language Models},
  author={Rai, Daking and Zhou, Yilun and Feng, Shi and Saparov, Abulhair and Yao, Ziyu},
  journal={arXiv preprint arXiv:2407.02646},
  year={2024}
}

@article{li2023starcoder,
  title={StarCoder: May the Source Be with You!},
  author={Li, Raymond and Allal, Loubna Ben and Zi, Yangtian and Muennighoff, Niklas and Kocetkov, Denis and Mou, Chenghao and Marone, Marc and Akiki, Christopher and Li, Jia and Chim, Jenny and others},
  journal={arXiv preprint arXiv:2305.06161},
  year={2023}
}

@inproceedings{wang2023codet5plus,
  title={CodeT5+: Open Code Large Language Models for Code Understanding and Generation},
  author={Wang, Yue and Le, Hung and Gotmare, Akhilesh Deepak and Bui, Nghi DQ and Li, Junnan and Hoi, Steven CH},
  booktitle={Proceedings of the 2023 Conference on Empirical Methods in Natural Language Processing},
  pages={1069--1088},
  year={2023}
}

@article{hui2024qwen25coder,
  title={Qwen2.5-Coder Technical Report},
  author={Hui, Binyuan and Yang, Jian and Cui, Zeyu and Yang, Jiaxi and Liu, Dayiheng and Zhang, Lei and Liu, Tianyu and Zhang, Jiajun and Yu, Bowen and Dang, Kai and others},
  journal={arXiv preprint arXiv:2409.12186},
  year={2024}
}

@article{team2024gemma2,
  title={Gemma 2: Improving Open Language Models at a Practical Size},
  author={Gemma Team and Riviere, Morgane and Pathak, Shreya and Sessa, Pier Giuseppe and Hardin, Cassidy and Bhupatiraju, Surya and Hussenot, Léonard and Mesnard, Thomas and Shaber, Bobak and Haber, Alexandre and others},
  journal={arXiv preprint arXiv:2408.00118},
  year={2024}
}

@article{ke2017lightgbm,
  title={Lightgbm: A highly efficient gradient boosting decision tree},
  author={Ke, Guolin and Meng, Qi and Finley, Thomas and Wang, Taifeng and Chen, Wei and Ma, Weidong and Ye, Qiwei and Liu, Tie-Yan},
  journal={Advances in neural information processing systems},
  volume={30},
  year={2017}
}
\bibliographystyle{icml2025}

\newpage
\appendix
\onecolumn
\section{Experimental Settings}
\label{app:exp_settings}

\subsection{Dataset Statistic}
\label{app:dataset_stat}
We report the final statistics of our curated multilingual MBPP incremental datasets in Table~\ref{tab:mbpp_dataset_stat}. Following our evaluation setup in Section~\ref{expe}, we use MBPP as the Python benchmark and construct semantically equivalent C++ and Java task sets by translating the same specifications, ensuring that the three datasets share comparable underlying algorithmic logic while differing only in surface syntax. Each JSONL instance corresponds to one labeled generation record with a binary correctness label (answer$\in\{0,1\}$), where $1$ denotes correct and $0$ denotes wrong. In total, our corpus contains 1,447 Python instances, 3,423 C++ instances, and 3,126 Java instances. The label distribution exhibits noticeable language-dependent skew: Python has the highest correctness ratio (81.6\% correct vs.\ 18.4\% incorrect), while C++ and Java are more challenging with substantially higher error rates (C++: 66.8\%/33.2\%; Java: 61.7\%/38.3\%). This imbalance motivates us to evaluate \method{} across languages and to study whether attribution-graph structural signatures remain predictive under varying error prevalence and programming syntax.

\begin{table}[h]
\centering
\begin{tabular}{lccc}
\toprule
\textbf{Dataset} & \textbf{Total Steps} & \textbf{\% Correct} & \textbf{\% Incorrect} \\
\midrule
C++ & 3{,}423 & 66.8\% & 33.2\% \\
Python & 1{,}447 & 81.6\% & 18.4\% \\
Java & 3{,}126 & 61.7\% & 38.3\% \\
\bottomrule
\end{tabular}
\caption{Final statistics of our curated MBPP incremental datasets, showing the number of instances and the distribution of correct/incorrect labels.}
\label{tab:mbpp_dataset_stat}
\end{table}

\subsection{Experimental Hyperparameter}
\label{app:hyperparam}
For attribution-graph extraction, we use the ReplacementModel instantiated from the base LLM, Google Gemma-2-2B-IT\footnote{\url{https://huggingface.co/google/gemma-2-2b-it}}, together with the corresponding GemmaScope transcoder\footnote{\url{https://huggingface.co/mntss/gemma-scope-transcoders}}, and run all tracing in bfloat16 precision on a single GPU. We use the Circuit-Tracer~\cite{circuit-tracer} library for attribution graph computation. For each input prompt, we construct an attribution graph with the following settings: we attribute at most 10 output logits and target a cumulative logit probability mass of 0.95. To control graph size and memory footprint, we cap the number of retained feature nodes to 8,192 and compute attributions in mini-batches of size 64. We disable parameter offloading and verbose debugging output. Following the preprocessing in our pipeline, we skip prompts longer than 550 characters to avoid excessively long contexts during tracing. Each resulting attribution graph is serialized in PyTorch format (.pt), and we additionally append a per-instance metadata record in JSONL format, including the task identifier, step index, graph path, and the corresponding binary correctness label.

\subsection{Detailed Features}
\label{app:detailed_features}
For each line-level attribution graph, we extract a fixed-length feature vector from the pruned subgraph to characterize both node composition and circuit topology. Concretely, we first prune the full attribution graph using an influence-based masking rule with a node-threshold of 0.8, which retains only the most influential nodes according to the cumulative influence scores returned by the pruning routine. Let the pruned node set contain (i) transcoder feature nodes, (ii) error nodes (one error node per layer-position pair), as well as (optionally) token and logit nodes if they survive pruning. We then compute three groups of features.

\emph{(1) High-level statistics.} We record (a) the total number of active features in the original graph (the length of the activation-value vector), (b) the number of pruned feature nodes, and (c) the number of pruned error nodes. To summarize output uncertainty, we additionally compute (d) the top-1 logit probability and (e) the entropy of the traced logit probability vector. These quantities capture the overall circuit scale and the model’s output confidence for the step.

\emph{(2) Aggregated node and activation statistics.} Using the cumulative influence scores restricted to the pruned nodes, we compute (a) the mean influence over all pruned nodes as a global saliency summary. For error nodes, we compute (b) the total error influence (sum over pruned error nodes) and (c) the mean error influence, serving as a proxy for how much of the computation is routed through unexplained residual components. For pruned feature nodes, we gather their activation values and compute (d) mean activation, (e) maximum activation, and (f) activation standard deviation. Finally, to capture where computation concentrates across depth, we construct a (g) layer-wise histogram of pruned feature counts: for each transformer layer, we count how many retained feature nodes originate from that layer. This produces a length-$L$ vector, where $L$ is the number of transformer layers.

\emph{(3) Topological and edge-based features of the pruned subgraph.} We form a directed weighted graph from the pruned adjacency matrix, where directed edges represent attributed influence and edge weights are inherited from the adjacency matrix. From edge weights, we compute (a) the sum, (b) mean, and (c) standard deviation of all edge weights, together with (d) the number of edges in the pruned subgraph and (e) the graph density. To capture structural fragmentation, we compute (f) the number of weakly connected components. We then compute centrality-based summaries: (g) mean degree centrality, (h) maximum degree centrality, (i) mean betweenness centrality (computed with edge weights), and (j) maximum betweenness centrality. To quantify global connectivity, we compute (k) the average shortest path length within the largest weakly connected component (set to $-1$ if undefined). Finally, we add a task-relevant routing signal: (l) the shortest-path distance from any retained input-token node to any retained logit node (set to $-1$ if no path exists), which measures how directly input evidence can flow to the output decision within the pruned circuit. For degenerate cases where the pruned graph has fewer than two nodes, we return a default feature vector with zeros and sentinel values (e.g., $-1$ for undefined shortest-path statistics) to preserve a consistent feature dimensionality across all examples.

\section{Prompt Template}
\label{app:prompt_templates}

\begin{table*}[t]
\centering
\small
\begin{tabular}{p{0.97\linewidth}}
\toprule
\textbf{Gemma-2-2B-IT Code Generation Prompt Template} \\
\midrule
You are a code assistant. \\
You must output ONLY valid code. \\
Do not include comments, markdown, explanations, or extra text. \\
Output only code. \\
\\
Task prompt: \\
\{TASK\_PROMPT\} \\
\bottomrule
\end{tabular}
\caption{Prompt template used for code generation. The final input is the concatenation of the fixed instruction block above and the task prompt, separated by two newline characters.}
\label{tab:prompt_codegen}
\end{table*}

\begin{table*}[t]
\centering
\small
\begin{tabular}{p{0.97\linewidth}}
\toprule
\textbf{Line-wise Code Correctness Checker Prompt Template} \\
\midrule
You are a code correctness checker. \\
Review the provided code line by line. For each line, output a score of 1 or 0 according to the following rules: \\
\\
1. If the line has a syntax error, output 0. \\
2. If the line is syntactically correct but violates the coding requirements described in the task, output 0 for that line and all subsequent lines. Once a line violates the requirements, all following lines must also be 0. \\
3. If the line is syntactically correct and meets the coding requirements so far, output 1. \\
\\
The code has exactly \{N\_LINES\} lines. Output only a list of exactly \{N\_LINES\} integers, e.g., [1, 1, 0, 1]. Do not include any explanation or extra text. \\
\\
4. Focus solely on whether the substantive code contains issues, ignoring parameter variable names, function names, and so on. \\
\\
Task Description: \\
--- \\
\{TASK\_DESCRIPTION\} \\
--- \\
\\
Code to Check: \\
--- \\
\{CODE\_TO\_CHECK\} \\
--- \\
\bottomrule
\end{tabular}
\caption{Prompt template used for line-wise correctness scoring. \{N\_LINES\} is computed as the number of lines in \{CODE\_TO\_CHECK\}; the model is instructed to output a list of exactly \{N\_LINES\} binary integers.}
\label{tab:prompt_linewise_checker}
\end{table*}

For code generation, we use a fixed instruction block that strictly constrains the model to output only valid code; this block is concatenated with the task prompt (two newline characters in between) to form the final input, as shown in Table~\ref{tab:prompt_codegen}. For correctness checking, we use a single user prompt that instructs a judge model to perform line-wise validation and return an integer list of length \{N\_LINES\}, where any violation at a given line forces that line and all subsequent lines to be scored as 0; we then parse the returned list as per-line correctness signals, as shown in Table~\ref{tab:prompt_linewise_checker}.

\section{Limitation}
The primary limitation of this approach is the computational overhead required to construct attribution graphs. Unlike black-box methods, our framework involves gradient-based backpropagation and feature projections, which increase time and memory complexity relative to standard inference. For extensive code generation tasks, these requirements may limit real-time deployment. Nevertheless, this bottleneck presents a significant opportunity for algorithmic optimization. Future research could focus on sparse attribution techniques that target only critical logical junctions or utilize low-rank approximations to reduce the cost of Jacobian computations. By transitioning from exhaustive circuit tracing to selective structural diagnostics, the computational burden can be substantially mitigated, paving the way for efficient, white-box verification in production environments.


\end{document}